\documentclass[3p,times,twocolumn]{elsarticle}
 \biboptions{comma,sort&compress,merge,mcite}
 \usepackage{hyperref}
   \usepackage{flushend}
\usepackage{graphicx}
\usepackage{amsmath,amssymb,amsbsy}
\usepackage{here}
\usepackage{ecrc}

\volume{00}

\firstpage{1}

\journalname{Nuclear and Particle Physics Proceedings}

\runauth{}

\jid{nppp}

\jnltitlelogo{Nuclear and Particle Physics Proceedings}

\usepackage{amssymb}

\usepackage[figuresright]{rotating}

\def\vec#1{\boldsymbol{#1}}
\newcommand{\SpSp}[2]{ \mbox{$\vec{\sigma }_{#1}.\vec{\sigma }_{#2}$}}
\newcommand{\lala}[2]{ \mbox{${\tilde{\lambda}_{#1}.\tilde{\lambda}_{#2}}$}}
\newcommand{\LLSS}[2]{\lala{#1}{#2}\,\SpSp{#1}{#2}}
\begin{document}

\begin{frontmatter}

\title{Exotic hadrons with heavy flavour or hidden flavour$^*$}
 \cortext[cor0]{Talk given at 18th International Conference in Quantum Chromodynamics (QCD 15,  30th anniversary),  29 june - 3 july 2015, Montpellier - FR}
 \author[label1]{Jean-Marc Richard}
\ead{j-m.richard@ipnl.in2p3.fr}
\address[label1]{Universit\'e de Lyon, Institut de Physique Nucl\'eaire de Lyon, IN2P3-CNRS-UCBL, \\
4, rue Enrico Fermi, Villeurbanne, France}
\pagestyle{myheadings}
\markright{ }
\begin{abstract}
The molecular picture and the constituent-quark model of exotic hadrons are reviewed, with application to the states recently discovered in the hidden-charm and hidden beauty sectors, and to other configurations.
\end{abstract}
\end{frontmatter}
\section{Introduction}
Since the discovery of the $X(3872)$ \cite{Choi:2003ue,Agashe:2014kda}, there has been a lot of experimental activity in the field of $XYZ$ mesons, and a flurry of theoretical papers.  The literature benefits from valuable reviews, for instance
%
and several talks at this Conference have given interesting updates of the experimental and theoretical studies. In this contribution, I shall make a few comments about the molecular model and the constituent-quark pictures. The field of mesons with hidden or naked  heavy flavour will be slightly extended, as to include some remarks on multi-baryons with strangeness (i.e., light hypernuclei), and the pentaquarks which have been announced \cite{Aaij:2015tga} in between this QCD15 conference and the deadline for its Proceedings. 

It has been early noticed that states with a variety of quarks, such as $(uuddss)$, can benefit from a coherent chromomagnetic attraction without offending the Pauli principle. Introducing charm or beauty is not just a variant. The mass of the $c$ or $b$ quarks gives more binding, and call for a subtle interplay of spin-independent and spin-dependent interactions, or, say, chromoelectric and chromomagnetic effects.

The possibilities offered by mixing light and heavy flavours was noticed rather early, for instance in \cite{Dover:1977jw} for multi-baryons, in \cite{Ader:1981db} for double-charm mesons, in
\cite{Gignoux:1987cn,*Lipkin:1987sk}%
\footnote{It is pleasure to thank Harry Lipkin for several enjoyable discussions and to renew the best wishes of the community at the occasion of his 94th birthday which occurred shortly before this Conference. Note that the two papers \cite{Gignoux:1987cn,*Lipkin:1987sk} were elaborated independently and are always cited together, except, surprisingly,  in a recent paper by the LHCb collaboration \cite{Aaij:2015tga}. 
  }%
for anticharmed baryons, etc., at a time where the experimental search was not easy. The new experimental facilities offer a considerable improvement and should give access to many configurations, even beyond hidden-charm and hidden-beauty.

Already for double-charm or double-beauty baryons, there is a subtle interplay between the slow relative motion of the heavy quarks, and the interaction of the flavoured core with the relativistic light quark \cite{Karliner:2014gca}.
\section{Molecules}
The idea is rather old that the Yukawa interaction is not restricted to nucleons. The physics of hyperons linked to nuclei is now very well documented. There remains, however, some uncertainty about configurations which are at the edge of binding. For years, the main guidance was provided by models based on meson exchanges and SU(3) flavour symmetry, carefully tuned to accommodate the sparse amount of date on hyperon-nucleon and hyperon-hyperon interaction. See, e.g., \cite{Nagels:2015dia}, and refs.\ there to the earlier literature. This approach is, however, challenged by models based on  chiral effective field theory \cite{Haidenbauer:2013oca} which give comparable scattering lengths but appreciably smaller effective range, and thus favours the binding of Borromean%
\footnote{A Borromean three-body system is bound while its two-body subsystems are unbound. For a Borromean $n$-body system with $n>3$, there is no path to build the system by adding the constituents one by one through a series of bound states.} %
states, as shown recently \cite{Garcilazo:2014mra,*Richard:2014pwa,*Ando:2015fsa}.

In the case of charm, besides hypernuclei \cite{Dover:1977jw}, there has been early speculations about the possibility of $D^{(*)}\bar D{}^{(*)}$ molecules \cite{Voloshin:1976ap}, in particular to account for some anomalies in the decay pattern of high-lying $\Psi$ states \cite{Bander:1975fb,*Rosenzweig:1975fq,*DeRujula:1976qd,*Minami:1978vf}, but this picture was abandoned and the anomalies eventually explained by the nodes in  the  wavefunction of the radially excited $(\bar c c)$ states \cite{LeYaouanc:1977ux,*LeYaouanc:1977gm,*Eichten:1979ms}. 

More recently, the possibility of molecules was revisited by Ericson and Karl \cite{Ericson:1993wy}, Manohar and Wise \cite{Manohar:1992nd}, and especially T\"ornqvist \cite{Tornqvist:1991ks,*Tornqvist:1993ng}, and the discovery of the $X(3872)$ \cite{Choi:2003ue} was considered as a success for this approach. Now, a more moderate point of view tends to prevail, where the $X(3872)$ is mainly a $(\bar c c)$ state, in which the higher Fock components play a more important role than in other quarkonium states \cite{Ferretti:2013faa}.

The scenario of the $X(3872)$ as a molecule, though very appealing, call for some warnings. The degeneracy or near degeneracy of isospin $I=0$ and $I=1$  is badly broken, as some of the leading terms of the interaction, such as $\pi$- or $\rho$-exchange are strongly isospin dependent. Historically, this isospin-dependence of nuclear forces made it difficult to describe the mesons as baryon-antibaryon molecules \cite{Ball:1965sa}.

Another problem is the risk of proliferation. Binding $D$ and $\bar D{}^*$  suggests the existence of similar meson-meson bound states with hidden beauty, and also of meson-baryon bound states, baryon-baryon bound states \cite{Julia-Diaz:2004rf,*Meguro:2011nr}, etc. In ordinary nuclear physics, the existence of repulsive forces at short distances restricts the number of bound states. For instance, the spin-singlet proton-proton and neutron-neutron systems are unbound though the pion-exchange is attractive, and the spin-triplet proton-neutron is rescued by the interplay of the s- and d-wave components \cite{Blatt:2019224}. In most other hadron-hadron systems, there is no reason to believe that there is a repulsive core in the interaction, and thus attractive long-range forces lead more easily to bound states. 
\section{Chromomagnetism}
Let us now consider a direct quark picture of tentative multiquark states. Jaffe \cite{Jaffe:1976yi} and many authors after him stressed the role of the short-range chromomagnetic interaction. Very schematically, once the other degrees of freedom are integrated out, the chromomagnetic Hamiltonian reads
\begin{equation}\label{eq:Hcm}
H_{\mathrm{CM}} =\sum_i m_i- \sum_{i,j} C_{ij}\,\LLSS{i}{j}~,
\end{equation}
where $C_{ij}$ is the expectation value of a short-range operator.  

In the sector of ordinary mesons and baryons, this Hamiltonian explains most of the observed hyperfine splittings, such as $J/\psi-\eta_c$. 

If applied to the the dibaryon configuration $H=(uuddss)$ with the assumption that the $C_{i,j}$ are the same for all pairs, and taken equal to their value for ordinary baryons, a binding of about $-150\,$MeV is obtained, below the $\Lambda\Lambda$ threshold. 

Note that this excess of attraction is rather remarkable. For instance, to describe the positronium molecule Ps$_2$ in terms of the Coulomb interaction $\sum a_{ij}/r_{ij}$, the same cumulated strength $\sum a_{ij}=-2$ is found for both Ps$_2$ and its threshold made of two isolated atoms, and  the binding of Ps$_2$ is due to a subtle deformation of its constituents, to favour the attractive terms. 

In subsequent studies of the $H$ dibaryon, it was shown that breaking SU(3) flavour symmetry and computing the $C_{ij}$  self-consistently from 6-body 
wavefunctions considerably reduce the attraction and suggest that the $H$ is unbound  \cite{Oka:1983ku,*Rosner:1985yh,*Karl:1987cg}.

More recently, the model \eqref{eq:Hcm} has been applied to the $X(3872)$ described as $(c\bar c q\bar q)$ with $J^P=1^+$ and $I=0$ \cite{Hogaasen:2005jv}. The parameters $C_{ij}$, and the effective quark masses $m_i$ (which include the chromoelectric energy) were taken from a fit of heavy hadrons. A remarkable state was found near 3872\,MeV which is a pure octet-octet of colour in the $(c\bar c)+(q\bar q)$ channel and thus is refrained from falling apart into a charmonium and an ordinary meson. The decay into a charmed and an anticharmed mesons is suppressed by the lack of phase space. This model predicts an isospin $I=1$ partner slightly above, with the same quantum numbers. Unfortunately, the data \cite{Agashe:2014kda} suggests that the $X(3872)$ and the neutral $X(3900)$ have opposite charge conjugations, and thus different structures. 
\section{More detailed quark model}
Standard multiquark calculations in the quark model rely on the Hamiltonian
\begin{multline}\label{eq:full-H}
  H=\sum_{i=1}^5 m_i+\sum_i\frac{\vec{p}_i^2}{2\,m_i}
  -\sum_{i < j}\lala{i}{j}\,V(r_{ij})\\
  {}-\frac{\gamma}{m_i\,m_j}\,\LLSS{i}{j}\,\tilde{\delta}{}_\mu^{(3)}~,
  \end{multline}
 of which the centre-of-mass motion is automatically removed if one uses translation-invariant wavefunctions in the variational calculation. The masses $m_i$ are modified as compared to \eqref{eq:Hcm} as the confining interaction is explicitly taken into account. The pairwise character of the chromoelectric term is questionable, and some alternative have been proposed \cite{Vijande:2011im,Vijande:2013qr,Cardoso:2013pta}. However, the above colour-octet-exchange ansatz can serve as a  guide for exploratory studies. 
 
 If the chromomagnetic term is omitted, i.e., $\gamma=0$, then the simple toy model \eqref{eq:full-H} looks very similar to the Hamiltonian governing atomic systems, with the colour algebra \lala{i}{j} replacing the Abelian product of charges $e_i\,e_j$, but leads to different results in the case of equal masses. In atomic physics, the positronium molecule Ps$_2$ is stable, by a small margin. If one solves \eqref{eq:full-H} with $\gamma=0$ and equal masses, no normalisable state is found below the threshold. In particular, the binding of $(cc\bar c\bar c)$ suggested by Llyod and Vary \cite{Lloyd:2003yc} is not confirmed in other calculations \cite{Ader:1981db,Carlson:1988hh,Barnea:2006sd}.
 
 For unequal masses, the patterns induced by symmetry breaking are similar in atomic and quark physics. The $(M^+M^+m^-m^-)$ molecule becomes more stable if the mass ratio $M/m$ increases. Similarly, some bound state occurs for $(QQ\bar q\bar q)$ if the mass ratio becomes large. For systems such a $(M^+m^+M^-m^-)$, the stability is lost if $M/m\gtrsim  2.2$ (or, of course $M/m\lesssim 1/2.2$), but some metastability can be envisaged with respect to the highest threshold $(M^+,m^-)+\mathrm{c.c.}$. In quark physics, the existence of $(Q\bar Qq\bar q)$ is also to be discussed with respect to $(Q\bar q)+\mathrm{c.c.}$ and relies on chromomagnetic ($\gamma\neq 0$) or long-range Yukawa interaction that is not included in \eqref{eq:full-H}.
 
 For $\gamma\neq0$, the spin-spin interaction is treated non-perturbatively, and thus the short-range correlation factor, $\langle \tilde{\delta}{}_\mu^{(3)} \rangle$, is computed consistently for each multiquark instead of being guessed from the hyperfine splittings of ordinary hadrons.
 
 In particular, solving $H$ in the $(c\bar c q\bar q)$ sector with $J^{PC}=1^{++}$ and $I=0$ confirms the results of the pure chromomagnetic model \cite{Hogaasen:2005jv}, namely a state is found near the $D\bar D{}^*$ threshold which is almost pure octet-octet of colour in the $(c\bar c)-(q\bar q)$ channel.
\section{Dibaryons}
The holly grail in this field would be a stable multiquark, with exotic quantum numbers and weak decay.  There is a persisting interest into mesons with double charm or beauty %
\cite{Ader:1981db,Carlson:1988hh,Janc:2004qn,*Carames:2011zz,*Hyodo:2012pm,*Bicudo:2015vta}, say $(QQ\bar q\bar q)$, which are very likely stable. One notices there that the collective configuration benefits from the chromoelectric interaction between the two heavy quarks and from the chromomagnetic interaction between the two light antiquarks, in the case of $J^P=1^+$, while the threshold $(Q\bar q)+(Q\bar q)$ is deprived of both. 

The question is whether the double-charm or double-beauty tetraquark is somewhat unique, or some other configurations exist, that combine astutely chromoelectric and chromomagnetic effects to reach a low mass below the fall-part threshold. A possibility is perhaps offered by dibaryons with two heavy quarks, say $D_{QQ}=(QQqqqq)$, where $Q$ stands for $c$ or $b$ and now $q$ denotes $u$, $d$ or $s$, but the game is seemingly more subtle than for the doubly-flavoured tetraquarks. On the chromoelectric side, $D_{QQ}$ benefits from the attraction between the two heavy quarks, and so does the threshold $T\!h_1=(QQq)+(qqq)$. As for the chromomagnetic interaction, in the limit of a large mass ratio $M(Q)/m(q)$, $D_{QQ}$
behaves similarly to the 1987-vintage pentaquark $(\bar Q qqqq)$ \cite{Gignoux:1987cn,*Lipkin:1987sk}, and gets a cumulated strength which is about twice that of $T\!h_1$, as does the second threshold $T\!h_2=(Qqq)+(Qqq)$.  So the dibaryon  benefits from both effects, while each threshold is restricted to a single one. The hope is that $D_{QQ}$ will eventually be stable, provided the variational wavefunction is flexible enough to incorporate the short- and long-range contributions. 

Another formulation could be the following. The dibaryon $D_{QQ}$ has two thresholds, and in empirical models such as \eqref{eq:full-H}, they are almost degenerate. Hence if one thinks of  a cluster model of the six-body system
\begin{multline}\label{eq:cluster}
 \psi(\vec r, \vec r', \dots)=\alpha(\vec r)\,|Qqq\rangle|Qqq\rangle\\
 {}+\beta(\vec r')|QQq\rangle|qqq\rangle~,
\end{multline}
the two components will interfere efficiently to lower the mass of the compound. In \eqref{eq:cluster}, $\vec r$ and $\vec r'$ denote the interbaryon separation  and the ellipses stand for the other Jacobi coordinates that are needed to describe the relative motion of the six quarks. The estimate of the energy of various $D_{QQ}$ states is currently under active study.
\section{Pentaquarks}
As already mentioned in the introduction, pentaquark states have been revealed in the $J/\psi + p$ mass spectrum of the decay $\Lambda_b\to J/\psi K^- p$ studied by the LHCb collaboration \cite{Aaij:2015tga}. 

In the early days of baryon physics, there have been speculations about possible resonances of the $KN$ interaction, the so-called $Z$-resonances \cite{Ross:1975da}. 

Some new activity in this sector was motivated by the development of colour chemistry \cite{Chan:1978nk} to account for the baryonium bound states and resonances. See, for instance, \cite{Aerts:1977rw,*deCrombrugghe:1978hi}. But the interest did not survive the death of baryonium.

The pentaquark with a single heavy constituent, $(\bar c qqqq)$, was looked at  by the E791 collaboration at Fermilab \cite{Aitala:1997ja,*Aitala:1999ij}, but the search was not conclusive. 

More recently, there were some speculations about a low-lying antidecuplet of light baryons. If translated in terms of the quark model, these states correspond to $(\bar q qqq)$, where, again, $d$ denotes $u$, $d$ or $s$. See, e.g.,~\cite{Praszalowicz:2005ps}, and  refs.\ there. Some experimentalists were curious enough to probe the idea, and found a tentative candidate \cite{Nakano:2003qx}, followed by  a gregarious sequence of positive signals. Unfortunately, as reviewed in \cite{Agashe:2014kda}, the light pentaquark was not confirmed in most forthcoming experiments, especially the ones with efficient (and expensive) detectors.

There is already a number of papers about the hidden-charm pentaquarks, and several studies are about to be published. In the constituent quark model, it has been noticed years ago that the chromomagnetic interaction acts favourably in some $(\bar Q Q qqq)$ configurations  \cite{Leandri:1989su}. It remains to perform a full five-body calculation to see whether some metastability emerges from the interplay of kinetic energy, confinement and chromomagnetism.  Experimentally, it would be interesting to study other decay channels of $\Lambda_b$. The paper by Moinester et al.\ remains a good guide \cite{Moinester:1995qi} for selecting the final states suited for pentaquarks.
\section{Outlook}
The idea of a rich spectroscopy of exotic or crypto-exotic states including both heavy and light quarks and antiquarks is now rather old, but, to be honest,  the hadrons found do not always fit the detailed predictions. One way out is perhaps to improve the treatment of the light quark dynamics. Within the Born-Oppenheimer method, one can combine an elaborate treatment of  the light degrees of freedom and the potential model for the heavy quarks or antiquarks \cite{Hasenfratz:1977dt,*Hasenfratz:1980jv,*Braaten:2014qka}. The quarkonium potential, $v(r)=3\,V(r)/16\sim -a/r +b\,r+c$, to match the notation of \eqref{eq:full-H}, is the ground-state energy of the gluon field surrounding a $Q\bar Q$ pair frozen at a separation $r$. Then the Schr\"odinger equation with $v(r)$ as input, supplemented by the spin-dependent corrections,   gives the radial and orbital excitations of charmonium. Now, if the gluon field is in its first excited state, another potential $v^*(r)$ is obtained, and the Schr\"odinger equation with $v^*(r)$ gives the so-called hybrid mesons. In the early papers, the potentials $v(r)$ and $v^*(r)$ were estimated empirically using a variant of the bag model. This approach has been significantly improved within lattice QCD \cite{Juge:1999ie}. The new development is that not only gluon excitations but also quark degrees of freedom are introduced \cite{Braaten:2013oba}. In particular, the $XYZ$ mesons are governed by an effective $c\bar c$ (or $b\bar b$) potential $\bar v(r)$ corresponding to the minimal energy of the gluons and light quarks surrounding  the heavy quarks of given separation $r$. The recent pentaquarks can be described within an effective $c\bar c$ potential minimising, for given $r$, the energy corresponding to three light quarks, the gluon field and some virtual pairs. Note that the Born-Oppenheimer approximation usually works very well, and can be probed in toy models. For instance, it can be checked that for a given interquark potential, a Born-Oppenheimer treatment of the double-charm baryons reproduces almost exactly the results form an accurate three-body calculation of $ccq$ \cite{Fleck:1989mb}.

The molecular picture is very appealing in identifying the main degrees of freedom in multiquark states, but remains restricted to states close tho the threshold. In nuclear physics, there is a repulsive core that refrains the nucleon from overlapping. Now if two hadrons carrying  heavy flavour feel an attractive long-range force, there constituents will unavoidably come close together and interact directly by exchange of gluons. Also, the simplest molecular configurations have been already listed, and the new ones require a study of detailed interplays between several hadron-hadron states, as noticed, e.g., in \cite{Roca:2015dva} for the recent pentaquarks.

The diquark model has been rather successful in hadron spectroscopy \cite{Anselmino:1992vg}, though the early contributions are sometimes forgotten. An interesting development was proposed to explain the baryonium states \cite{Chan:1978nk}, with a diquark and an antidiquark seprated by an orbital-momentum barrier. The recent applications to states without relative orbital momentum between the clusters is of course somewhat more speculative, but perhaps this approach \cite{Maiani:2014ola} will reveal some mechanisms of the interaction among quarks. 

Experimentally, this research remains delicate, as only a few flavour configurations are accessible. The negative results on the search for  double-charm baryons illustrate how difficult it is to identify hadrons with several heavy quarks.

\section*{Acknowledgements}
I would like to thank  Stephan Narison for organising this beautiful series of conferences, Emiko Hiyama, Alfredo Valcarce, Javier Vijande and Marek Karliner for very useful discussions just before QCD15, and Marina Nielsen and Fernando Navarra for further discussions during our daily probes of the local flavours in liquid form.

This paper is dedicated to the memory of my colleague and friend Claude Gignoux, a pioneer in this physics.

\end{document}